\documentclass[aps,prb,twocolumn,floatfix,nofootinbib,superscriptaddress]{revtex4}
\usepackage{amsfonts}
\usepackage{amsmath}
\usepackage{amssymb}
\usepackage{eucal}
\usepackage{psfrag}
\usepackage{tabularx}
\usepackage{graphicx,psfrag,subfigure}
\usepackage{amsbsy}

\def\be{\begin{equation}}
\def\ee{\end{equation}}
\def\ba{\begin{eqnarray}}
\def\ea{\end{eqnarray}}

\def\LSCO{La$_{2-x}$Sr$_x$CuO$_4$}

\def\C60{A$_x$C$_{60}$}

\def\HgCu3{HgCa$_2$Cu$_3$O$_{8+y}$}
\def\HgCu4{HgBa$_2$Ca$_3$Cu$_4$O$_{10+y}$}
\def\TlCu{Tl$_2$Ba$_2$CuO$_{6+\delta}$}
\def\TlCu3{Tl$_2$Ba$_2$Ca$_2$Cu$_3$O$_{10+y}$}
\def\TlCu4{Tl$_2$Ba$_2$Ca$_3$Cu$_4$O$_{12+y}$}

\def\BiCu3{Bi$_2$Sr$_2$Ca$_{2}$Cu$_3$O$_y$}

\def\C60{A$_x$C$_{60}$}

\begin{document}

\title{ Charge $4e$ superconductivity from pair density wave order in
certain high temperature superconductors}
\author{ Erez Berg}
\affiliation{Department of Physics, Stanford University, Stanford, California 94305-4060,
USA}
\author{ Eduardo Fradkin}
\affiliation{Department of Physics, University of Illinois at Urbana-Champaign, Urbana,
Illinois 61801-3080, USA}
\author{ Steven A. Kivelson}
\affiliation{Department of Physics, Stanford University, Stanford, California 94305-4060,
USA}
\date{\today }

\maketitle

\textbf{A number of spectacular experimental anomalies\cite
{li-2007,fujita-2005} have recently been discovered in certain
cuprates, notably {La$_{2-x}$Ba$_x$CuO$_4$} and {La$_{1.6-x}$Nd$%
_{0.4}$Sr$_x$CuO$_{4}$}, which exhibit unidirectional spin and
charge order (known as ``stripe order''). We have recently
proposed to interpret these observations as evidence for a novel
``striped superconducting'' state, in which the superconducting
order parameter is modulated in space, such that its average is
precisely zero. Here, we show that thermal melting of the striped
superconducting state can lead to a number of unusual phases, of
which the most novel is a charge $4e$ superconducting state, with
a corresponding fractional flux quantum $hc/4e$. These are
never-before observed states of matter, and ones, moreover, that
cannot arise from the conventional Bardeen-Cooper-Schrieffer (BCS)
mechanism. Thus, direct confirmation of their existence, even in a
small subset of the cuprates, could have much broader implications
for our understanding of high temperature superconductivity. We
propose experiments to observe fractional flux quantization, which
thereby
could 
confirm the existence of these states.}

It is widely accepted that the mechanism of superconductivity in
the cuprate high temperature superconductors is different than in
conventional superconductors, and that the ``normal state'' above
$T_c$ is anything but normal. In contrast, the low energy low
temperature properties of the superconducting state appears to be
largely accounted for by the conventional BCS theory, albeit with
a d-wave order parameter and a much suppressed superfluid density.
In short, the perception is that the superconducting (SC) phase is
simple while the ``normal'' phase is not. However, recent
experiments \cite{li-2007,fujita-2005} in the ``stripe ordered''
materials, in which a novel form of highly two-dimensional partial
SC order was found, suggest that the SC phases may have
unconventional aspects. We have
interpreted\cite{berg-2007,berg-2008a,berg-2009} these anomalies
as evidence of a new type of order, ``striped superconductivity''
or unidirectional pair density wave (PDW) order. Subsequent experiments\cite%
{tranquada08,basov08} have provided additional indirect evidence
in favor of this interpretation.

\begin{figure}[hbt]
\begin{center}
\includegraphics[width=0.48\textwidth]{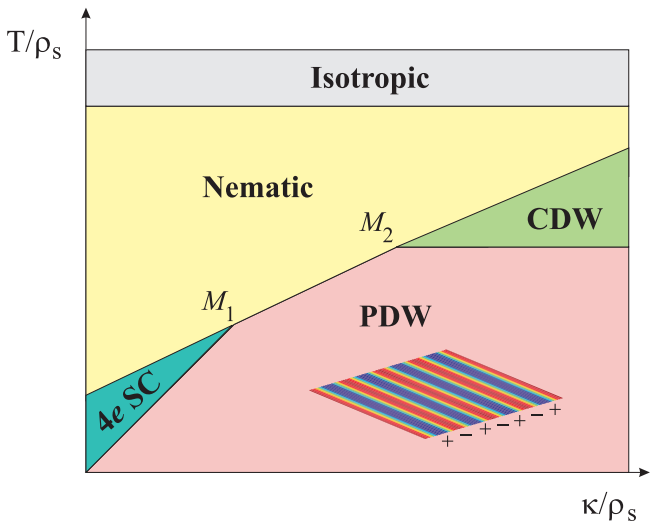}
\end{center}
\caption{Schematic phase diagram: KT transitions between the PDW, CDW
(stripe), nematic, and charge $4e$ SC, and the (Ising) nematic-isotropic
transition. $M_1$ and $M_2$ are multi-critical points (see text). PDW, CDW,
and charge $4e$ SC orders have QLRO in their respective phases, and are
short ranged in the nematic phase.}
\label{fig:phase-diagram}
\end{figure}

In the present paper, we study the thermal melting of a striped
superconducting groundstate by the proliferation of topological
defects leads to a complex phase diagram, shown in
Fig.\ref{fig:phase-diagram}, with several interesting phases, of
which the most novel is a charge $4e$ superconducting state. The
existence of such a state can be directly established
experimentally by observing magnetic flux quantization with period
$hc/4e$, half that of the usual superconducting
flux quantum, in a variety of experimental geometries, as shown in Fig.\ref%
{fig:squid}. Moreover, it is likely that the charge 4e SC, resulting from
partial melting the striped SC, has a finite density of states for gapless
quasiparticles, reflecting the fact that it does not arise from the Bose
condensation of literal four electron bound states.

In related developments, Agterberg and Tsunetsugu\cite
{agterberg08,agterberg-2009} and Radzihovsky and Vishwanath, \cite
{Radzihovsky-2008} have discussed related phenomena in connection
with Larkin-Ovchinnikov (LO) phases in, respectively, the heavy
fermion material CeCoIn$_5$ at high fields and in partially
spin-polarized ultra-cold atomic gases. The states we discuss here
differ from the LO states in that there is no explicit
time-reversal symmetry breaking. Also the structure of the striped
SC reflects the discrete point group symmetry of the underlying
(cuprate) lattice.

A striped superconductor is a unidirectional PDW state in which a spin
singlet superconducting order parameter, $\Delta(\mathbf{r})$ oscillates in
space with a wave vector $\mathbf{Q}$,\cite{commensuration} (shown
schematically in Fig. \ref{fig:phase-diagram}):
\begin{equation}
\Delta(\mathbf{r})=\Delta_{\mathbf{Q}}(\mathbf{r}) \, e^{i \mathbf{Q} \cdot
\mathbf{r}}+ \Delta_{-\mathbf{Q}}(\mathbf{r}) \, e^{-i \mathbf{Q} \cdot
\mathbf{r}}.  \label{eq:op-pdw}
\end{equation}
Of course, there will always be higher harmonics of the order parameter at
wave vectors $n\vec Q$, but in the ordered state, these are slaved to the
fundamentals and hence are not independent dynamical degrees of freedom.
Generally, therefore, we will not treat them explicitly. However, two
subsidiary order parameters play a special role in the thermal melting:
Charge density wave (CDW) order, $\rho(\mathbf{r})=\rho_0+\rho_{\mathbf{K}}(%
\mathbf{r}) \, e^{i \mathbf{K} \cdot \mathbf{r}}+ \rho^\star_{\mathbf{K}}(%
\mathbf{r}) \, e^{-i \mathbf{K} \cdot \mathbf{r}}$, appears as a second
harmonic of the fundamental ordering, with wave vector $\mathbf{K}=2\mathbf{Q%
}$, where $\rho_{\mathbf{K}}(\mathbf{r}) \propto \Delta^\star_{-\mathbf{Q}}(%
\mathbf{r}) \Delta_{\mathbf{Q}}(\mathbf{r})$. Similarly, charge $4e$ SC
order, represented by the complex scalar field $\Delta_{4e}(\mathbf{r})
\propto \Delta_{-\mathbf{Q}}(\mathbf{r}) \Delta_{\mathbf{Q}}(\mathbf{r})$,
occurs parasitically along with the fundamental striped SC order.

Although the stripe ordered superconductor need not be associated with a
non-zero magnetization, it is in many ways similar to a LO state.\cite%
{Larkin-1964} As such it has a $U(1) \times U(1)$ symmetry corresponding to
the independent uniform shifts of the phases of the complex components of
the order parameter:
\begin{eqnarray}
\Delta_{\pm \mathbf{Q}}(\mathbf{r}) && \to e^{i\theta_{\pm \mathbf{Q}}}
\Delta_{\pm \mathbf{Q}}(\mathbf{r}) =e^{i(\theta\pm \phi)} \Delta_{\pm
\mathbf{Q}}(\mathbf{r}), \\
\rho_{\mathbf{K}}(\mathbf{r}) && \to e^{i2\phi} \rho_{\mathbf{K}}(\mathbf{r}%
)\ \ \mathrm{with } \ \ \phi\equiv (\theta_{\mathbf{Q}}-\theta_{-\mathbf{Q}%
})/2  \notag \\
\Delta_{4e}(\mathbf{r}) &&\to e^{i 2\theta}\Delta_{4e}(\mathbf{r}) \ \
\mathrm{with } \ \ \theta\equiv (\theta_{\mathbf{Q}}+\theta_{-\mathbf{Q}})/ 2
\notag  \label{eq:symmetries}
\end{eqnarray}
Much as other stripe electronic liquid crystal states, the PDW also
typically breaks the point group symmetry; to be concrete, we will consider
the case of a tetragonal crystal, in which the choice of direction for the
unidirectional order breaks 
the $C_{4v}$ point group symmetry down to $C_2$. In analogy with classical
liquid crystals, we refer to this as nematic ordering, although since the
symmetry breaking is in fact, discrete, it is an ``Ising nematic.'' This is
the minimal set of broken symmetries associated with the striped
superconductor. For simplicity, we will consider only the interplay of
charge stripe and SC stripe orders, and ignore the striped spin order, even
though all three are intertwined where striped superconductivity is
conjectured to occur in the cuprates.\cite{berg-2007,berg-2008a,berg-2009}

We consider a system which has a striped superconducting groundstate, and
consider the manner in which thermal fluctuations gradually restore all the
broken symmetries as the temperature, $T$, increases. Deep in the PDW phase,
other than in a vortex core, we can ignore fluctuations in the magnitude of
the order parameter, and write $\Delta_{\pm \mathbf{Q}}(\mathbf{r}%
)=\Delta_{SC}\, \exp\big\{ i\left(\theta(\mathbf{r})\pm \phi (\mathbf{r}%
)\right)\big\}$, $\rho_{\mathbf{K}}(\mathbf{r})=\rho_{K} \exp\left[i2\phi(%
\mathbf{r})\right]$, and $\Delta_{4e}(\mathbf{r})=\Delta_{4e}\exp\left[%
i2\theta(\mathbf{r})\right]$. Because the phase of the CDW order is subject
to pinning by quenched disorder, the PDW is considerably more fragile than a
uniform SC state, and quenched randomness easily leads to an XY
superconducting glass phase. (Unless stated otherwise, below we ignore the
effects of disorder.) Moreover, the striped superconductor state admits
novel topological excitations, which play a central role in the thermal
melting of the groundstate.

Deep in the striped SC phase, treating the system as 2D, and taking $\vec Q$
in the $x$ direction, the effective Hamiltonian for the low energy thermal
fluctuations is
\begin{equation}
\mathcal{H}[\theta,\varphi]= \frac{\rho_s}{2} \left|\boldsymbol{D_s}%
\theta\right|^2+ \frac{\kappa}{2} \left(\boldsymbol{D_c} \varphi\right)^2,
\label{eq:london}
\end{equation}
$\boldsymbol{D_s} = \alpha_s[-i\hbar \partial_x -(2e/c) A_x]\hat x
+ \alpha_s^{-1}[-i\hbar \partial_y -(2e/c) A_y]\hat y$,
$\boldsymbol{D_c} = \alpha_c[-i\hbar \partial_x ]\hat x +
\alpha_c^{-1}[-i\hbar \partial_y]\hat y $, $\rho_s$ and $\kappa$
are, respectively, the superfluid stiffness and the CDW elastic
constant, and $\alpha_s$ and $\alpha_c$ are the corresponding
(finite) anisotropies.\cite{continuum} We have shown, explicitly,
the coupling to an external gauge field, but henceforth, unless
otherwise specified, we will take $\mathbf{A}=\mathbf{0}$
\cite{comment-A}. The correlation functions of the PDW, CDW and
charge $4e$ SC order parameters in this phase are
\begin{eqnarray}
&& \langle \Delta_{\pm \boldsymbol{Q}}(\mathbf{r}) \Delta_{\pm \boldsymbol{Q}%
}(\mathbf{r}^\prime)^*\rangle \propto K_s(\mathbf{r} -\mathbf{r}^\prime) K_c(%
\mathbf{r} -\mathbf{r}^\prime)  \notag \\
&& \langle \rho_{ 2\boldsymbol{Q}}(\mathbf{r}) \rho_{2 \boldsymbol{Q}}(%
\mathbf{r}^\prime)^*\rangle \propto [K_c(\mathbf{r} -\mathbf{r}^\prime)]^4
\notag \\
&& \langle \Delta_{4e}(\mathbf{r}) \Delta_{4e}(\mathbf{r}^\prime)^*\rangle
\propto [K_s(\mathbf{r} -\mathbf{r}^\prime) ]^4  \label{eq:qlro}
\end{eqnarray}
where
\begin{eqnarray}
&& K_s(\mathbf{r}) \sim\big[(x\alpha_s)^2+(\alpha_s^{-1} y)^2\big]^{-\frac{%
\eta_s}{2}}  \notag \\
&& K_c(\mathbf{r}) \sim \big[(x\alpha_c)^2+(\alpha_c^{-1} y)^2\big]^{-\frac{%
\eta_c}{2}}  \notag \\
&& \eta_s= 2\pi T/\rho_s \ \ \ \eta_c=2\pi T/\kappa.
\end{eqnarray}

Since the order parameters $\Delta_{\pm \mathbf{Q}}(\mathbf{r})$ must be
separately single valued, their phase fields must be invariant under the
transformations $\theta_{\pm \mathbf{Q}}(\mathbf{r}) \to \theta_{\pm \mathbf{%
Q}}(\mathbf{r}) +2\pi m_{\pm \mathbf{Q}}$, where $m_{\pm \mathbf{Q}}$ are
integers. Correspondingly, the fields $\theta(\mathbf{r})$ and $\varphi(%
\mathbf{r})$ must obey the conditions $\theta(\mathbf{r}) \to \theta(\mathbf{%
r})+\pi (m_{ \mathbf{Q}}+ m_{- \mathbf{Q}})$, $\varphi(\mathbf{r}) \to
\varphi(\mathbf{r}) + \pi (m_{ \mathbf{Q}}-m_{-\mathbf{Q}})$. The integers $%
m_{\pm \mathbf{Q}}$ then classify the topological excitations supported by
the PDW state: vortices with topological charge $q_s=(m_{ \mathbf{Q}}+ m_{-
\mathbf{Q}})/2$ and dislocations with topological charge $q_c=(m_{ \mathbf{Q}%
}- m_{- \mathbf{Q}})/2$. We have three types of topological excitations $%
(q_s,q_c)$:

\begin{enumerate}
\item Full vortices with $q_s =\pm 1$ and $q_c=0$.

\item Double dislocations, with $q_c=\pm 1$ 
and $q_s=0$.

\item Half-vortices, $q_s=\pm 1/2$, bound to single dislocations, $q_c =\pm
1/2$.
\end{enumerate}

Much as in the case of the well understood theory of the Kosterlitz-Thouless
(KT) phase transition,\cite{kosterlitz73,nelson-1979,young-1979} the
thermodynamic behavior of this system is also represented by a generalized
(neutral) Coulomb gas of vortices $\{q_s\}$ and dislocations $\{ q_c\}$,
both with logarithmic interactions, where the strength of the interactions
between vortices (and anti-vortices), both fractional and integral, is
controlled by the superfluid density $\rho_s$ and the interaction between
dislocations (and anti-dislocations) is controlled by the CDW stiffness $%
\kappa$. Let us denote by $g_{({q_s,q_c})}$ the fugacity of a topological
excitation with topological charges $(q_s,q_c)$. In the dilute limit, in
which the fugacities are small, the partition function of the generalized
vector Coulomb gas\cite{young-1979} can be represented by a sine-Gordon type
effective field theory (for a review see Refs.\cite{Nienhuis1987}), that in
this case requires two fields $\tilde{\theta}$ and $\tilde{\varphi}$, the
\emph{dual} of the superconducting phase $\theta$ and the CDW phase $\varphi$%
. The effective Hamiltonian density of the 2D field theory, dual to the
degrees of freedom of Eq.\eqref{eq:london}, is
\begin{eqnarray}
&&\!\!\!\!\!\!\!\!\mathcal{H}_{\mathrm{dual}}[\tilde{\theta},\tilde{\varphi}%
]= \frac{T}{2\rho_s} \left(\boldsymbol{D}_s \tilde{\theta}\right)^2+ \frac{T%
}{2\kappa} \left(\boldsymbol{D}_c \tilde{\varphi}\right)^2  \notag \\
&& \!\!\!\!\!\!\!\! -g_{({1,0})} \cos(2\pi \tilde{\theta}) - g_{({0,1})}
\cos(2\pi \tilde{\varphi})  \notag \\
&& \!\!\!\!\!\!\!\!- 2g_{(\frac{1}{2},\frac{1}{2})} \cos (\pi \tilde{\theta}%
)\cos(\pi \tilde{\varphi})  \label{eq:SG1}
\end{eqnarray}

There are three distinct pathways for the thermal melting of the PDW state%
\cite{spin} by condensation of its topological excitations, as shown in the
schematic phase diagram in Fig.\ref{fig:phase-diagram}.
In constructing this figure, the location of the phase boundaries were
estimated by computing the scaling dimensions of the cosine operators at the
non-interacting fixed point, $\Delta_{q_s,q_c}=\frac{\pi}{T} (q_s^2
\rho_s+q_c^2 \kappa)$, and then identifying the lines, $\Delta_{q_s,q_c} = 2$%
, at which each of the cosine operators in Eq.\eqref{eq:SG1} first becomes
relevant. Alternatively, a mean-field phase diagram can be obtained by
treating Eq.\eqref{eq:SG1} in a self-consistent phonon approximation. More
sophisticated RG treatments of a Coulomb gas system with precisely the same
\textit{formal} structure as the above have been discussed previously for a
model of the thermal melting of a charge-spin stripe state,\cite{kruger-2002}
and the context of spinor $S=1$ Bose condensates.\cite{podolsky-2007} All
approaches yield the same topology of the phase diagram as the one shown in
Fig.\ref{fig:phase-diagram}, although there are some unresolved differences
concerning the shape of the various phase boundaries and the nature of the
multicritical points, $M_1$ and $M_2$. Moreover, since the specific problems
being addressed in \cite{podolsky-2007} and \cite{kruger-2002} were
different than those being considered here, the interpretation and the
physics of the phases was different.

\paragraph{PDW Phase:}

At low temperatures, all three cosine operators are irrelevant and the
topological excitations are uncondensed. The system is in a PDW state with
quasi-long range order (QLRO), as given in Eq. \eqref{eq:qlro}.

\paragraph{Charge $4e$ SC Phase:}

This phase is accessed when the 
double dislocations proliferate, \textit{i.e.\/} when the operator $%
\cos(2\pi \tilde{\varphi})$ becomes relevant, Hence, the PDW-Charge $4e$SC
phase boundary is the line $\Delta_{0,1}=4\pi \kappa/T=2$, or $\frac{\kappa}{%
\rho_s}=2 \pi \frac{T}{\rho_s}$ (see Fig.\ref{fig:phase-diagram}.) In the
charge $4e$ SC phase, the field $\tilde{\varphi}$ is pinned at integer
values $n_{\varphi} \in \mathbb{Z}$, and has massive (gapped) fluctuations.
In this phase, the dislocations of the CDW are screened, and the dual the
CDW phase field $\varphi$ has wild fluctuations, leading to exponentially
decaying correlations of both the PDW and CDW order parameters. In contrast,
the field $\tilde{\theta}$ remains gapless, and the correlation function of
the charge $4e$ SC have power-law correlations, with the same exponent $%
\eta_{4e}$. The phase transition from the charge $4e$ SC to the disordered
non-superconducting nematic higher temperature phase proceeds through the
subsequent unbinding and proliferation of vortices with fractional (half)
topological charge.

\paragraph{Stripe (CDW) Phase:}

This phase is accessed through the proliferation of SC vortices with integer
topological charge when the operator $\cos(2\pi \tilde{\theta}_+)$ becomes
marginal, $\Delta_{1,0}=\pi \rho_s/T=2$. There is CDW QLRO, with correlation
functions that decay with a power law with an exponent $\eta_{CDW}$, and no
SC order of any type. In this phase the field $\tilde{\theta}$ is pinned at
the integer values $n_\theta \in \mathbb{Z}$, and its fluctuations are
massive. The phase transition from this phase to the normal (nematic) phase
occurs by the unbinding of \emph{single dislocations} whose fractional
vortex charge is screened in this phase.

\paragraph{Direct PDW-nematic normal state phase transition:}

This transition proceeds through the liberation of half-vortex-integer
dislocation composite excitations, \textit{i.e.\/} when the topological
excitations $(\pm \frac{1}{2},\pm \frac{1}{2})$ condense. This phase
boundary is the $M_1 M_2$ line shown in Fig. \ref{fig:phase-diagram},
determined by the condition $\Delta_{\pm \frac{1}{2},\pm \frac{1}{2}}=2$.

\paragraph{Multicritical Points:}

The schematic phase diagram of Fig.\ref{fig:phase-diagram} has two
multi-critical points, $M_1$ and $M_2$, whose existence follows from the
topology of the phase diagram. From the structure of Eq. \ref{eq:SG1}, it
also follows that there exist two special critical points with an emergent
higher symmetry where two operators become marginal simultaneously: $(0,1)$
and $(\frac{1}{2},\pm \frac{1}{2})$ at $M_1^\prime$, and $(1,0)$ and $(\frac{%
1}{2},\pm \frac{1}{2})$ at $M_2^\prime$. At these points, the general $%
U(1)\times U(1)$ symmetry of the model is enlarged to an $SU(3)_1$ symmetry,%
\cite{kondev-henley96} and in their vicinity the correlation length exhibits
KT behavior with a modified exponent.\cite{young-1979,kruger-2002} The
simple scaling arguments used to construct Fig.\ref{fig:phase-diagram} and
the RG analysis in \cite{kruger-2002} suggest that the high symmetry and
multicritical points are one and the same, $M_j=M_j^\prime$, while the
self-consistent phonon calculation and the RG treatment of \cite%
{podolsky-2007} suggest that they are distinct, with $M_1^\prime$ and $%
M_2^\prime$ lying part way along the segment of the phase boundary between $%
M_1$ and $M_2$.

While the present discussion of thermal melting was confined to 2D, it can
be readily extended to the case of a 3D layered material with sufficiently
weak inter-layer coupling. Some aspects of this extension depend on details
of the interlayer geometry\cite{berg-2009}. For instance, depending on
whether or not the inter-layer coupling frustrates the stripe ordering, the
low temperature state may or may not have a gentle, time-reversal symmetry
breaking superconducting spiral superimposed on the basic single-plane
stripe order. But, in general, all of the phases indicated in Fig.\ref%
{fig:phase-diagram} persist in 3D, with the QLRO replaced by true long-range
order, and a crossover, very near the phase boundaries, to 3D criticality.

At $T=0$ 
the power-law correlations of the PDW phase are replaced by true long range
order and gapless collective (Goldstone) modes. A simple mean-field treatment%
\cite{dror,berg-2009} of the spectrum of Bogoliubov quasiparticles in this
phase reveals that, in common with a CDW, it generically has an only
partially gapped Fermi surface, and hence a finite density of zero energy
states. In principle, it should also be possible for the PDW phase to
undergo \emph{quantum melting}, possibly also by a quantum analog of the
vortex unbinding mechanism described above.\cite{quantum-melting} This is a
more complex problem than the thermal melting we have discussed here. In
particular, the \emph{quantum} phase transitions are affected by the
existence of gapless quasiparticles. Hence, the structure and quantum
dynamics of the vortices (both integer and fractional) is expected to be
generally damped and anisotropic, which affects the physics of quantum
melting.

\begin{figure}[t]
\begin{center}
\subfigure[]{\includegraphics[width=0.2\textwidth]{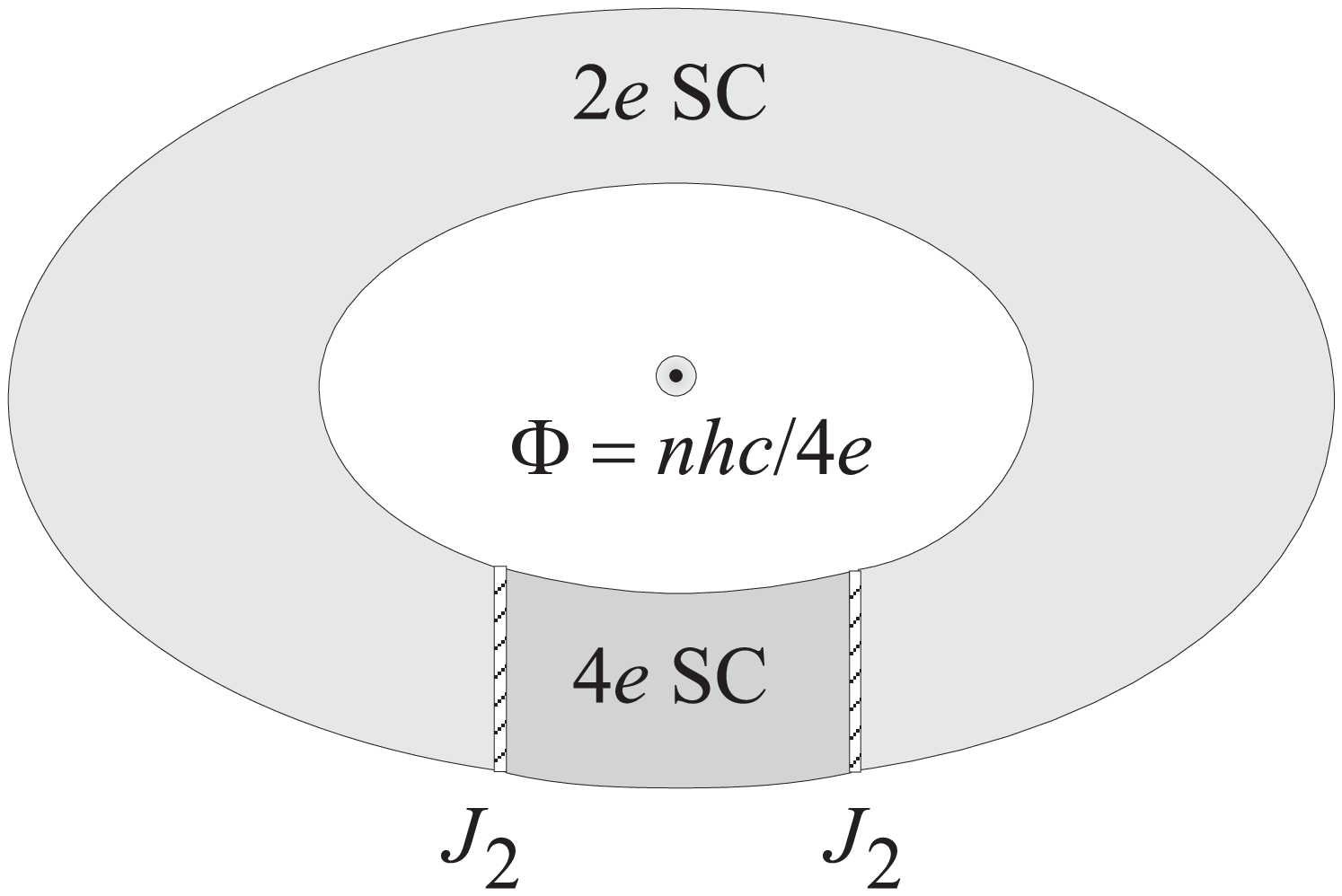}} %
\subfigure[]{\includegraphics[width=0.2\textwidth]{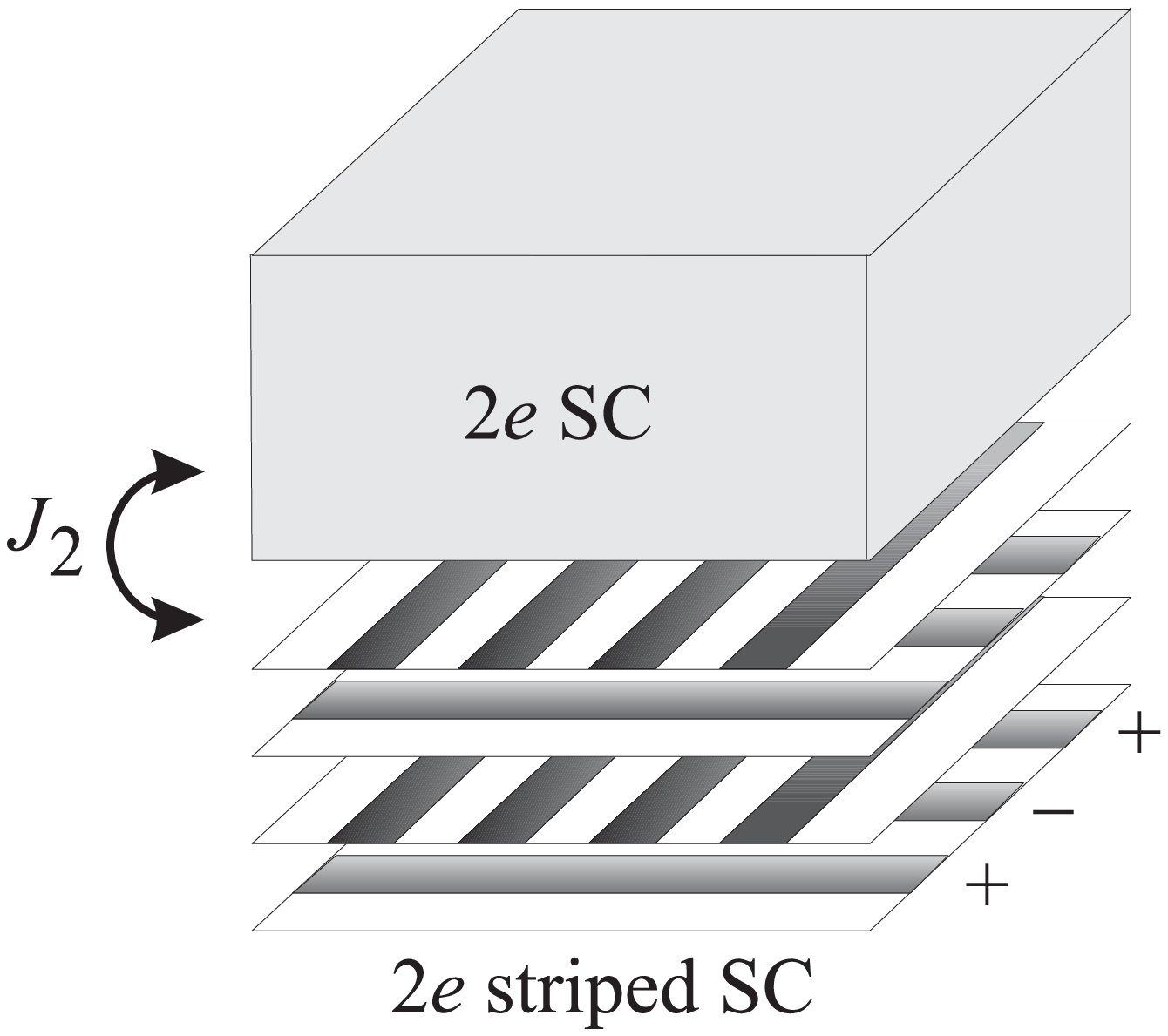}}
\end{center}
\caption{Two schematic phase-sensitive experiments: a) a SQUID loop with a
charge 4e SC, and b) a charge 2e striped SC-charge 2e uniform SC Josephson
junction (see text).}
\label{fig:squid}
\end{figure}
There are many unusual physical consequences of the nature of the PDW phase
and of its thermally melted daughter phases. We have previously discussed%
\cite{berg-2007,berg-2008a,berg-2009} some remarkable bulk features of a
fully ordered PDW phase, including the possibility of dynamical layer
decoupling (in a 3D layered material), an anomalous sensitivity to quenched
disorder, the existence (in principle testable by STM and scanning SQUID) of
pinned half-vortices associated with each disorder induced dislocation, and
a tendency to formation of a superconducting glass phase which likely
spontaneously breaks time reversal symmetry. Here we focus on ``phase
sensitive'' measurements, which have not yet been explored, and offer the
possibility of directly establishing the existence of either the fully
ordered phase or its uniform, translationally invariant charge 4e SC
descendant.

Clearly, a superconducting SQUID loop, made of a charge 4e SC, will exhibit
all the familiar features of a SQUID, but with a distinct half-flux quantum,
$hc/4e$ replacing the usual supeconducting flux quantum, $hc/2e$. From a
practical point of view, however, it is undoubtedly easier to fabricate a
SQUID loop in which only a single link consists of a charge 4e SC, as shown
schematically in Fig.\ref{fig:squid}a. Such a SQUID will exhibit the same
half-flux quantum as if it were an entire loop of charge 4e SC. In addition
to the practical advantage, such a geometry will detect PDW related phases
under a wider range of circumstances. For instance, in the disordered phase
near criticality, so long as the width of the putative charge 4e link is not
large compared to the SC coherence length, it will simply act as a Josephson
weak link in a charge 4e SC SQUID. Moreover, under many circumstances, even
if the link consists of a PDW state, the SQUID will still exhibit $hc/4e$
periodicity.

To see this, consider a 
Josephson junction between a striped SC and an ordinary SC, as shown
schematically in Fig.\ref{fig:squid}b. The dependence of the free energy on
the phase difference, $\Delta \theta $, between the two SC can be expanded
as
\begin{equation}
F(\Delta \theta )=S\sum_{n=1}^{\infty }J_{n}\cos [n\Delta \theta ]\text{,}
\end{equation}%
where $S$ is the junction area, and $J_{n}$ involves tunneling processes of $%
2n$ electrons, 
so that generally $J_{n}$ decays exponentially with increasing $n$. For the
geometry shown in Fig. 2b, which is representative of the putative striped
SC phase in {La$_{2-x}$Ba$_{x}$CuO$_{4}$} near $x=1/8$, $J_{1}$ (and indeed,
all $J_{2n+1}$) vanish identically, since a $\pi $ phase change of the
striped SC is equivalent to a translation by half a period. As a result, the
(resistively shunted) Josephson coupling between the two SC is dominated by
the intrinsically smaller higher order coupling, $J_{2}$.\cite{half-flux}
Not only does that mean that a striped SC in the place of the charge 4e SC
link in Fig.\ref{fig:squid}a, will act in the same way, it also means that
all the standard characteristics of a single Josephson junction, including
Josephson oscillations and Shapiro steps, will occur with twice the usual
frequencies.

$J_{2}$ is proportional to $J^{2}$, where $J$ is the
\textquotedblleft bare\textquotedblright\ (microscopic) Josephson
coupling. A quantitative estimate of $J_{2}$ is difficult, since
it depends exponentially on microscopic parameters. However,
assuming that the coherence length $\xi $ is smaller than the
period of the striped superconductor, an estimate based on an
effective $x-y$ model (similar to that of Ref. \cite{berg-2009},
but adapted for a c-axis junction between a striped and a uniform
superconductor) gives that $J_{2}\sim J^{2}/\left\vert J^{\prime
}\right\vert $, where $J^{\prime }$ is the inter-stripe Josephson
coupling in the striped superconductor. Thus the energy
denominator is the smallest energy scale of the single-plane
problem. Moreover, since the total coupling energy is proportional
to the area of the junction, it should be possible to use a large
enough junction such that $SJ_{2}$ is substantial (i.e., much
larger than $k_{B}T$).

It should be possible to perform the proposed experiments using small
crystals of {La$_{2-x}$Ba$_x$CuO$_4$} with $x=1/8$, where there is already
strong circumstantial evidence of the existence of a striped SC phase. The
observation either of $hc/4e$ flux quantization in a SQUID of the sort shown
in Fig. 2a, or a charge $4e$ in the Josephson relation in a Josephson
junction of the sort shown in Fig. 2b, would constitute dramatic and direct
evidence of the existence of these exotic superconducting phases.

E. Berg, berez@stanford.edu, is the corresponding author.
\begin{acknowledgments}
We thank Dimitri Basov, Tony Leggett, Daniel Podolsky, Leo Radzihovsky, Doug
Scalapino, John Tranquada, and Dale Van Harlingen for discussions. This work
was supported in part by the National Science Foundation, under grants DMR
0758462 (E.F.) and DMR 0531196 (S.A.K.), and by the Office of Science, U.S.
Department of Energy under Contracts DE-FG02-91ER45439 through the Frederick
Seitz Materials Research Laboratory at the University of Illinois (E.F.),
DE-FG02-06ER46287 through the Geballe Laboratory of Advanced Materials at
Stanford University (S.A.K. and E.B.).
\end{acknowledgments}

All three authors contributed equally to all parts of this work. The authors declare that they have no
competing financial interests.


\begin{thebibliography}{10}
\expandafter\ifx\csname url\endcsname\relax
  \def\url#1{\texttt{#1}}\fi
\expandafter\ifx\csname urlprefix\endcsname\relax\def\urlprefix{URL }\fi
\providecommand{\bibinfo}[2]{#2}
\providecommand{\eprint}[2][]{\url{#2}}

\bibitem{li-2007}
\bibinfo{author}{Li, Q.}, \bibinfo{author}{H\"ucker, M.}, \bibinfo{author}{Gu,
  G.~D.}, \bibinfo{author}{Tsvelik, A.~M.} \& \bibinfo{author}{Tranquada,
  J.~M.}
\newblock \bibinfo{title}{{Two-Dimensional Superconducting Fluctuations in
  Stripe-Ordered La$_{1.875}$Ba$_{0.125}$CuO$_4$}}.
\newblock \emph{\bibinfo{journal}{Phys.\ Rev.\ Lett.}}
  \textbf{\bibinfo{volume}{99}}, \bibinfo{pages}{067001}
  (\bibinfo{year}{2007}).

\bibitem{fujita-2005}
\bibinfo{author}{Fujita, K.}, \bibinfo{author}{Noda, T.},
  \bibinfo{author}{Kojima, K.~M.}, \bibinfo{author}{Eisaki, H.} \&
  \bibinfo{author}{Uchida, S.}
\newblock \bibinfo{title}{{Effect of disorder outside the CuO$_{2}$ planes on
  $T_{c}$ of copper oxide superconductors}}.
\newblock \emph{\bibinfo{journal}{Phys. Rev. Lett.}}
  \textbf{\bibinfo{volume}{95}}, \bibinfo{pages}{097006}
  (\bibinfo{year}{2005}).

\bibitem{berg-2007}
\bibinfo{author}{Berg, E.} \emph{et~al.}
\newblock \bibinfo{title}{{Dynamical layer decoupling in a stripe-ordered high
  $T_c$ superconductor}}.
\newblock \emph{\bibinfo{journal}{Phys.\ Rev.\ Lett.}}
  \textbf{\bibinfo{volume}{99}}, \bibinfo{pages}{127003}
  (\bibinfo{year}{2007}).

\bibitem{berg-2008a}
\bibinfo{author}{Berg, E.}, \bibinfo{author}{Fradkin, E.} \&
  \bibinfo{author}{Kivelson, S.~A.}
\newblock \bibinfo{title}{{Theory of the Striped Superconductor}}.
\newblock \emph{\bibinfo{journal}{Phys. Rev. B}} \textbf{\bibinfo{volume}{79}},
  \bibinfo{pages}{064515} (\bibinfo{year}{2009}).

\bibitem{berg-2009}
\bibinfo{author}{Berg, E.}, \bibinfo{author}{Fradkin, E.},
  \bibinfo{author}{Kivelson, S.~A.} \& \bibinfo{author}{Tranquada, J.~M.}
\newblock \bibinfo{title}{{Striped superconductors: How the cuprates intertwine
  spin, charge and superconducting orders}} (\bibinfo{year}{2009}).
\newblock \bibinfo{note}{(unpublished)}, \eprint{arXiv:0901.4826}.

\bibitem{tranquada08}
\bibinfo{author}{Tranquada, J.~M.} \emph{et~al.}
\newblock \bibinfo{title}{{Evidence for unusual superconducting correlations
  coexisting with stripe order in La$_{1.875}$Ba$_{0.125}$CuO$_4$}}.
\newblock \emph{\bibinfo{journal}{Phys. Rev. B}} \textbf{\bibinfo{volume}{78}},
  \bibinfo{pages}{174529} (\bibinfo{year}{2008}).

\bibitem{basov08}
\bibinfo{author}{{A. A. Schafgans, A. D. LaForge, S. V. Dordevic, M. M.
  Qazilbash, S. Komiya, Y. Ando, and D. N. Basov}}.
\newblock \bibinfo{title}{{Magnetic-field-induced spin order quenches Josephson
  coupling in {\LSCO}}} (\bibinfo{year}{2008}).
\newblock \bibinfo{note}{(unpublished)}.

\bibitem{agterberg08}
\bibinfo{author}{Agterberg, D.~F.} \& \bibinfo{author}{Tsunetsugu, H.}
\newblock \bibinfo{title}{{Dislocations and vortices in pair-density-wave
  superconductors}}.
\newblock \emph{\bibinfo{journal}{Nature Phys.}} \textbf{\bibinfo{volume}{4}},
  \bibinfo{pages}{639} (\bibinfo{year}{2008}).

\bibitem{agterberg-2009}
\bibinfo{author}{Agterberg, D.~F.}, \bibinfo{author}{Sigrist, M.} \&
  \bibinfo{author}{Tsunetsugu, H.}
\newblock \bibinfo{title}{{Order parameter and vortices in the superconducting
  {$Q$}-phase of {CeCoIn$_5$}}} (\bibinfo{year}{2009}).
\newblock \bibinfo{note}{(unpublished)}, \eprint{arXiv:0902.0843}.

\bibitem{Radzihovsky-2008}
\bibinfo{author}{Radzihovsky, L.} \& \bibinfo{author}{Vishwanath, A.}
\newblock \bibinfo{title}{{Quantum liquid crystals in imbalanced Fermi gas:
  fluctuations and fractional vortices in Larkin-Ovchinnikov states}}
  (\bibinfo{year}{2008}).
\newblock \bibinfo{note}{(unpublished)}, \eprint{arXiv:0812.3945}.

\bibitem{commensuration}
\bibinfo{note}{Here we consider the case of incommensurate PDW order (or with
  high order commensuration)}.

\bibitem{Larkin-1964}
\bibinfo{author}{Larkin, A.~I.} \& \bibinfo{author}{Ovchinnikov, Y.~N.}
\newblock \bibinfo{title}{{Nonuniform state of superconductors}}.
\newblock \emph{\bibinfo{journal}{Zh. Eksp. Teor. Fiz.}}
  \textbf{\bibinfo{volume}{47}}, \bibinfo{pages}{1136} (\bibinfo{year}{1964}).
\newblock \bibinfo{note}{(Sov. Phys. JETP. \textbf{20}, 762 (1965))}.

\bibitem{continuum}
\bibinfo{note}{{In the continuum the charge anisotropy $\alpha_c$ diverges
  \cite{Radzihovsky-2008}}}.

\bibitem{comment-A}
\bibinfo{note}{This is justified, provided that the penetration depth is long
  enough.}

\bibitem{kosterlitz73}
\bibinfo{author}{Kosterlitz, J.~M.} \& \bibinfo{author}{Thouless, D.~J.}
\newblock \bibinfo{title}{{Ordering, metastability and phase transitions in
  two-dimensional systems}}.
\newblock \emph{\bibinfo{journal}{J. Phys. C}} \textbf{\bibinfo{volume}{6}},
  \bibinfo{pages}{1181} (\bibinfo{year}{1973}).

\bibitem{nelson-1979}
\bibinfo{author}{Nelson, D.~R.} \& \bibinfo{author}{Halperin, B.~I.}
\newblock \bibinfo{title}{{Dislocation-mediated melting in two dimensions}}.
\newblock \emph{\bibinfo{journal}{Phys. Rev. B}} \textbf{\bibinfo{volume}{19}},
  \bibinfo{pages}{2457} (\bibinfo{year}{1979}).

\bibitem{young-1979}
\bibinfo{author}{Young, A.~P.}
\newblock \bibinfo{title}{Melting and the vector coulomg gas in two
  dimensions}.
\newblock \emph{\bibinfo{journal}{Phys. Rev. B}} \textbf{\bibinfo{volume}{19}},
  \bibinfo{pages}{1855} (\bibinfo{year}{1979}).

\bibitem{Nienhuis1987}
\bibinfo{author}{Nienhuis, B.}
\newblock \bibinfo{title}{{Coulomb Gas Formulations of Two-dimensional Phase
  Transitions}}.
\newblock In \bibinfo{editor}{Domb, C.} \& \bibinfo{editor}{Lebowitz, J.}
  (eds.) \emph{\bibinfo{booktitle}{Phase Transitions and Critical Phenomena}},
  vol.~\bibinfo{volume}{11}, \bibinfo{pages}{1} (\bibinfo{publisher}{Academic
  Press}, \bibinfo{address}{London}, \bibinfo{year}{1987}).

\bibitem{spin}
\bibinfo{note}{The physics is more complex if the (intertwined) spin stripe
  order is also considered\cite{berg-2007} depending on whether the full
  $SU(2)$ spin symmetry remains intact (in which case there is no long range
  spin order in 2D for $T>0$), or if the magnetic anisotropy is XY-like or
  Ising-like.}

\bibitem{kruger-2002}
\bibinfo{author}{Kr\"uger, F.} \& \bibinfo{author}{Scheidl, S.}
\newblock \bibinfo{title}{Nonuniversal ordering of spin and charge in stripe
  phases}.
\newblock \emph{\bibinfo{journal}{Phys. Rev. Lett.}}
  \textbf{\bibinfo{volume}{89}}, \bibinfo{pages}{095701}
  (\bibinfo{year}{2002}).

\bibitem{podolsky-2007}
\bibinfo{author}{Podolsky, D.}, \bibinfo{author}{Chandrasekharan, S.} \&
  \bibinfo{author}{Vishwanath, A.}
\newblock \bibinfo{title}{{Phase Transitions of $S = 1$ Spinor Condensates in
  an Optical Lattice}} (\bibinfo{year}{2007}).
\newblock \bibinfo{note}{(unpublished)}, \eprint{arXiv:0707.0695v2}.

\bibitem{kondev-henley96}
\bibinfo{author}{Kondev, J.} \& \bibinfo{author}{Henley, C.~L.}
\newblock \bibinfo{title}{{Kac-Moody symmetries of critical ground states}}.
\newblock \emph{\bibinfo{journal}{Nucl. Phys. B}}
  \textbf{\bibinfo{volume}{464}}, \bibinfo{pages}{540} (\bibinfo{year}{1996}).

\bibitem{dror}
\bibinfo{author}{Baruch, S.} \& \bibinfo{author}{Orgad, D.}
\newblock \bibinfo{title}{Spectral signatures of modulated d-wave
  superconducting phases}.
\newblock \emph{\bibinfo{journal}{Phys. Rev. B.}}
  \textbf{\bibinfo{volume}{77}}, \bibinfo{pages}{174502}
  (\bibinfo{year}{2008}).

\bibitem{quantum-melting}
\bibinfo{note}{{The quantum melting of a charge-spin stripe state (without SC
  order) was discussed qualitatively in Ref.\cite{Nussinov-2002}.}}

\bibitem{half-flux}
\bibinfo{note}{{An analogous effect has been seen in Josephson junctions
  between two crystals of a $d$-wave superconductor with relative crystalline
  orientation chosen so that $J_1=0$\cite{schneider-2004}.}}

\bibitem{Nussinov-2002}
\bibinfo{author}{Nussinov, Z.} \& \bibinfo{author}{Zaanen, J.}
\newblock \bibinfo{title}{{Stripe fractionalization I: the generation of Ising
  local symmetry}}.
\newblock \emph{\bibinfo{journal}{Journal de Physique IV (Proceedings)}}
  \textbf{\bibinfo{volume}{12}}, \bibinfo{pages}{9} (\bibinfo{year}{2002}).

\bibitem{schneider-2004}
\bibinfo{author}{Schneider, C.~W.} \emph{et~al.}
\newblock \bibinfo{title}{{Half-h/2e critical current Oscillations of SQUIDs}}.
\newblock \emph{\bibinfo{journal}{Europhys. Lett.}}
  \textbf{\bibinfo{volume}{68}}, \bibinfo{pages}{86} (\bibinfo{year}{2004}).

\end{thebibliography}

\end{document}